\documentclass[sigconf]{acmart}

\acmConference[MobiHoc '26]{The 27th International Symposium on Theory, Algorithmic Foundations, and Protocol Design for Mobile Networks and Mobile Computing}{November 23--26, 2026}{Tokyo, Japan}
\acmYear{2026}
\copyrightyear{2026}
\setcopyright{acmlicensed}
\usepackage{graphicx}
\usepackage{booktabs}
\usepackage{algorithm}
\usepackage{algorithmic}
\usepackage{amsmath}
\usepackage{xspace}

\makeatletter
\g@addto@macro\UrlBigBreaks{\do\@}
\makeatother

\newcommand{\sysname}{\textsc{Skynet}\xspace}

\title[\sysname: Workflow-Level Anomaly Detection for Agentic AI]{\sysname: Workflow-Level Anomaly Detection for Agentic AI via Semantic and Structural Modeling}

\author{Chaoyu Zhang}
\authornote{Both authors contributed equally to this research.}
\affiliation{%
  \institution{Virginia Tech}
  \city{Arlington}
  \state{VA}
  \country{USA}
}
\email{chaoyu@vt.edu}

\author{Hexuan Yu}
\authornotemark[1]
\affiliation{%
  \institution{Virginia Tech}
  \city{Arlington}
  \state{VA}
  \country{USA}
}
\email{hexuanyu@vt.edu}

\author{Heng Jin}
\affiliation{%
  \institution{Virginia Tech}
  \city{Arlington}
  \state{VA}
  \country{USA}
}
\email{hengj@vt.edu}

\author{Shanghao Shi}
\affiliation{%
  \institution{WashU}
  \city{St.\ Louis}
  \state{MO}
  \country{USA}
}
\email{shanghao@wustl.edu}

\author{Ning Zhang}
\affiliation{%
  \institution{WashU}
  \city{St.\ Louis}
  \state{MO}
  \country{USA}
}
\email{zhang.ning@wustl.edu}

\author{Yi Shi}
\affiliation{%
  \institution{Virginia Tech}
  \city{Arlington}
  \state{VA}
  \country{USA}
}
\email{yshi@vt.edu}

\author{Yulia R. Gel}
\affiliation{%
  \institution{Virginia Tech}
  \city{Blacksburg}
  \state{VA}
  \country{USA}
}
\email{ygl@vt.edu}

\author{Y. Thomas Hou}
\affiliation{%
  \institution{Virginia Tech}
  \city{Blacksburg}
  \state{VA}
  \country{USA}
}
\email{hou@vt.edu}

\author{Wenjing Lou}
\affiliation{%
  \institution{Virginia Tech}
  \city{Arlington}
  \state{VA}
  \country{USA}
}
\email{wjlou@vt.edu}



\begin{document}

\begin{abstract}
Agentic AI systems execute complex tasks through long-horizon workflows of planning, tool use, and multi-agent coordination. Task failures in these systems often originate from a single step, such as an injected prompt or a flawed plan, and are then amplified through downstream dependencies as the corrupted step propagates across many subsequent agents and tool calls. Existing defenses either target a specific class of attacks or failures, or inspect individual prompts and steps in isolation. Both leave the global dependency structure of a workflow unexamined, and miss the inconsistencies that only emerge when the execution is viewed as a whole. We argue that anomaly detection for agentic AI must reason at the workflow level, where global execution structure exposes signals that local checks cannot see. We present \sysname, a principled workflow-level anomaly detection framework that turns observed multi-agent execution into directed workflow graphs and scores them against learned benign behavior. \sysname jointly models the semantic execution context and the structural organization of inter-agent delegation, tool invocation, and data-flow dependencies, and trains only on benign workflows. Because training never sees attacks or failures, this design naturally extends to zero-day detection: any execution that violates benign workflow regularities surfaces as off-manifold geometry under a single decision rule. We evaluate \sysname on three public agentic safety and failure benchmarks. It sustains high recall together with a sub-$1\%$ false positive rate, with per-workflow and per-step latencies low enough for online monitoring of agentic AI runtimes.
\end{abstract}

\maketitle

\section{Introduction}

Modern agentic AI stacks such as OpenAI's ChatGPT Agent~\cite{openai2024chatgptagent}, Anthropic's orchestrator-worker patterns~\cite{anthropic2024effectiveagents}, Microsoft's AutoGen~\cite{wu2024autogen}, and Google's Agent-to-Agent (A2A) protocol~\cite{google2025a2a} compose specialized large language model (LLM) agents that plan, invoke external tools through the Model Context Protocol (MCP)~\cite{anthropic_mcp_2024, openai_agents_python_mcp}, delegate sub-tasks, and exchange artifacts over long horizons~\cite{yao2022react, schick2023toolformer, wang2023voyager, wu2024autogen, yang2024swe, wang2024survey}. The system-level behavior of these stacks is best characterized by the \emph{workflows} they realize, which are ordered, dependency-rich execution structures over planning steps, tool calls, inter-agent messages, and system events, together with the temporal and causal links that determine how each step shapes the rest of the execution.

\begin{figure}[t]
  \centering
  \includegraphics[width=0.95\columnwidth]{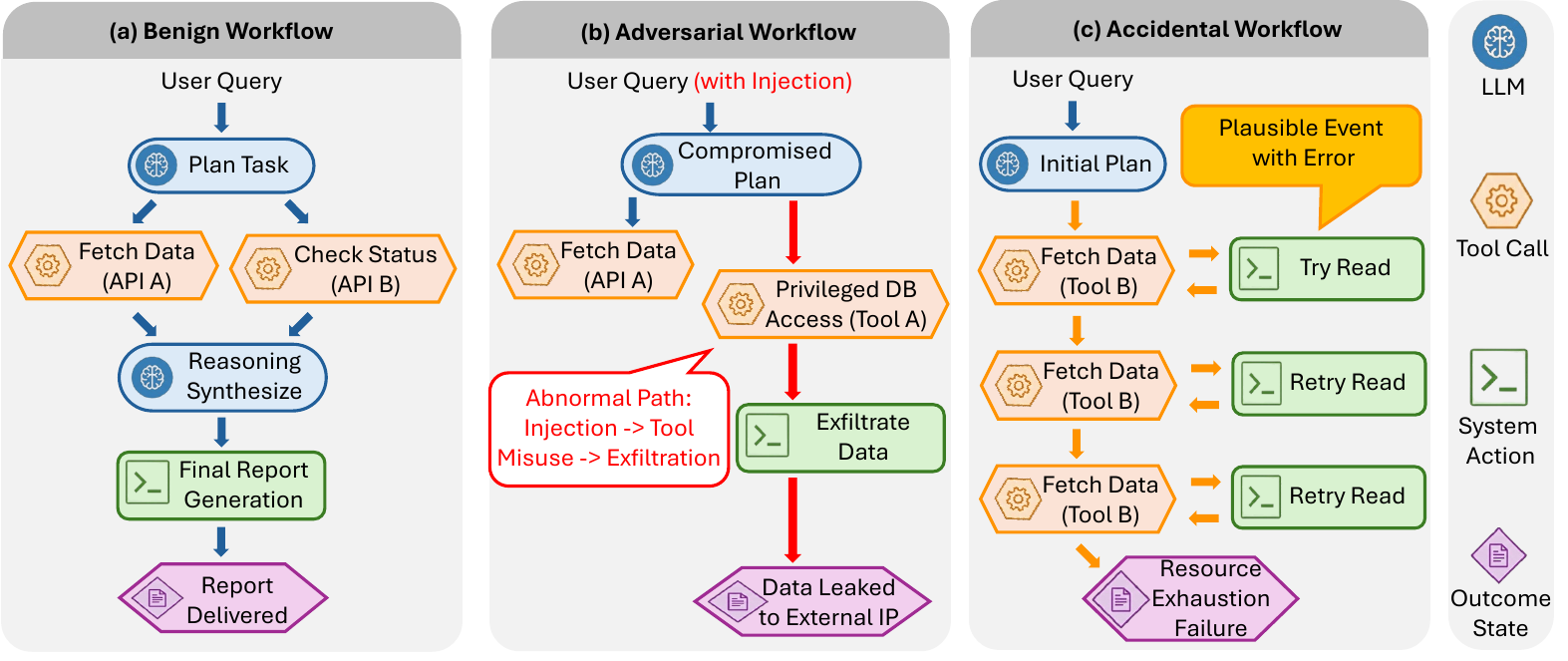}
  \caption{\textbf{Agentic AI execution workflows under normal and abnormal conditions.} (\textbf{a}) Benign workflow with coherent planning, tool use, and execution. (\textbf{b}) Adversarial manipulation, where injected steps redirect execution. (\textbf{c}) Intrinsic failure, where non-adversarial errors accumulate.}
  \Description{Three side-by-side agentic AI execution workflows drawn as directed graphs: a benign workflow, an adversarially manipulated workflow with injected steps, and an intrinsically failing workflow with accumulated errors.}
  \label{fig:workflow}
  \vspace{-0.5cm}
\end{figure}

This shift from single-call inference to long-horizon agentic execution introduces a qualitatively new \emph{risk surface} that subsumes both unintentional failures and adversarial threats under a single operational concern. The threats we observe on this surface come from two complementary sources. \emph{Intrinsic workflow failures} arise from the system itself, including flawed planning, ambiguous user instructions, hallucinated intermediates, and breakdowns in tool use or coordination~\cite{zhou2025guardian, debenedetti2025defeating, li2025safeflow, yu2026minim}. \emph{Adversarial workflow manipulations} arise from intentional interference such as prompt injection, tool-description injection, privilege escalation, or laundering of malicious intent across multiple agents~\cite{shi2026thinktwice, debenedetti2024agentdojo, zhan2024injecagent, ma2026safety, he2025red}. Despite these very different origins, both classes typically begin from a single perturbed step, such as a benign-yet-erroneous user request, a flawed plan, or a misjudged observation on one side, or a malicious instruction or tampered tool description on the other. Once such a step is produced, the workflow keeps consuming and propagating it: downstream agents condition on it, tools receive it, and later decisions are made on top of it, so the original perturbation is amplified across the rest of the execution. As a result, both surface as abnormal execution trajectories that diverge from intended task behavior (Figure~\ref{fig:workflow}).

Existing approaches mitigate this risk by attaching defenses to specific layers of the agentic stack. One line restricts how agents handle untrusted inputs~\cite{debenedetti2025defeating}, enforces execution-time invariants~\cite{wang2025agentarmor}, or applies transactional protocols where formal specifications are available~\cite{li2025safeflow}. Another line contains compromised components through sandboxing~\cite{wu2025isolategpt}, mediated tool access~\cite{li2025ace}, and rule-based action verification~\cite{chenshieldagent}. A third line matches execution traces against patterns associated with known attack and failure classes~\cite{liu2025traceaegis, li2025drift, zhou2025guardian}. Each of these is anchored to a specific class of problems or operates at the granularity of an individual prompt, action, or trace fragment, leaving the global dependency structure of a workflow only indirectly examined. As production agentic systems continuously add new tools, agents, and task domains, this leaves a long tail of cross-step anomalies that no single specification or pattern catches.

An important research question raised by this gap is therefore: \emph{at what granularity should anomalies in agentic AI be detected?} Network intrusion detection went through the same transition. Early packet-based detectors flagged each packet in isolation and missed attacks that only became visible across a long session, which motivated record-based intrusion detection that aggregates packets into flows and reasons about them as a whole. Agentic AI is in the same position. Inspecting a single prompt, action, or trace fragment throws away the dependency structure that an injected or flawed step uses to spread, and the most informative signals are non-local, including role-inconsistent delegations, atypical tool-call pathways, and abnormal artifact-to-agent influence. A monitor for agentic AI must therefore reason over whole workflows, model both their semantics and their dependency structure, and remain agnostic to which particular attack or failure mode is in play.

We realize this view in \sysname, a principled workflow-level anomaly detection framework for agentic AI. \sysname plugs into the observation layer of modern agentic runtimes by consuming the structured execution records they already expose, and consolidates them into a directed workflow graph that captures planning, tool invocations, inter-agent delegation, data flow, and system events with their temporal and causal dependencies. On top of this representation, \sysname learns from benign workflows alone, jointly modeling \emph{what} happens (semantic execution context) and \emph{how} effects propagate (dependency structure), with training views built around the structural regularities of benign agentic execution rather than from random graph perturbation. The detector is therefore aligned with the benign workflow manifold, and exposes both threat classes as off-manifold geometry under a single decision rule. Because training never sees attacks or failures, this design naturally extends to \emph{zero-day} detection: previously unseen attacks and novel failure modes that violate benign workflow regularities surface through the same off-manifold geometry, with no need to anticipate them.

We evaluate \sysname on three public agentic benchmarks targeting complementary failure regimes: Agent-SafetyBench~\cite{zhang2024agent} (adversarial manipulations), AgentErrorBench~\cite{zhu2025llmagenterror} (intrinsic execution failures), and ATBench~\cite{li2026atbench} (broad safety-in-the-wild). Training uses only benign trajectories from \textsc{ETO}~\cite{song2024eto} and never observes adversarial or failing workflows.


\section{System Model, Threat Model, and Problem}
\label{sec:prelim}

\subsection{System Model}
\label{sec:system-model}

\textbf{Target agentic AI systems.}
We target modern multi-agent AI systems that execute complex user intents through structured, long-horizon task execution: orchestrator--worker architectures~\cite{anthropic2024effectiveagents}, conversational multi-agent frameworks such as Microsoft AutoGen~\cite{wu2024autogen} and OpenAI's ChatGPT Agent~\cite{openai2024chatgptagent}, agent-to-agent interoperable ecosystems built on Google's A2A protocol~\cite{google2025a2a}, and ReAct-style planners~\cite{yao2022react, wang2023voyager} that invoke tools through the Model Context Protocol (MCP)~\cite{anthropic_mcp_2024, openai_agents_python_mcp}. A common pattern recurs across these stacks: a planner (or orchestrator) agent decomposes a user task into subtasks, delegates them to specialized worker agents, and composes intermediate results through tool calls and inter-agent messages. The planner naturally acts as an \emph{observer} of the execution it initiates, since the runtime already exposes structured traces of what each agent did, in what order, with which tools, and with what intermediate results. \sysname plugs directly into this observation layer as a passive consumer of execution records that already exist for telemetry, debugging, and compliance; it does not modify agent logic, tool backends, inter-agent protocols, or LLM weights.

\textbf{Observed execution records.}
Our system model focuses on \emph{observed execution workflows} rather than the internal planning states of individual agents. We assume access to four classes of records standard across modern agentic platforms: (i) \emph{planning and reasoning traces} (ReAct-style thought/action/observation tuples, orchestrator plans, intermediate scratchpads); (ii) \emph{tool invocations} (MCP-style calls with typed inputs/outputs, calling agent identity, status); (iii) \emph{inter-agent messages} through protocols such as A2A or AutoGen's conversation primitives, including delegation requests, replies, and shared artifacts; and (iv) \emph{system-level events} (retries, errors, sandbox decisions, timeouts). When a planner delegates a task, it monitors the resulting execution and collects these records within its visibility scope; nested delegations record evidence locally and return it on completion, so workflow evidence is propagated bottom-up and consolidated at the root planner.

\textbf{Workflow construction.}
Given the consolidated record of an execution, \sysname builds a directed workflow graph (Sec.~\ref{sec:methodology}) in a runtime-agnostic way. Nodes correspond to observed execution steps annotated with structured attributes that summarize \emph{what the step was}, including actor role, action type, tool/endpoint identifier, and status. Typed edges capture \emph{how steps relate}: temporal precedence, delegation between agents, data-flow dependencies, and cross-agent communication. This construction is intentionally protocol-agnostic and treats platform-specific record formats as instantiations of the same abstraction; for example, an AutoGen conversation, an A2A-mediated interaction, a ChatGPT Agent trace, or a general MCP-tool workflow each yields a graph populated under the same node and edge schema once the corresponding records are parsed through a thin adapter. As a result, a detector trained on one deployment can be reused across heterogeneous agentic systems without retraining on platform-specific schemas, and the framework can be extended to additional runtimes by supplying a new adapter rather than redesigning the representation.

\textbf{Trust boundary.}
We assume that workflow measurements collected within an organizational trust boundary are trustworthy, supported by standard system-security mechanisms on the host infrastructure. When execution is delegated to external services or agents outside this boundary, we assume the availability of remote attestation mechanisms for exchanging \emph{verifiable workflow-level evidence}~\cite{zhang2021recfaRA6, wang2023ariRA7, ammar2024bridgingRA8, ammar2024RAsok}. The design of secure measurement and attestation infrastructures is orthogonal to our contribution.

\subsection{Threat Model}
\label{sec:threat}

\textbf{Adversary capabilities and limitations.}
We consider adversaries operating \emph{within} the trusted measurement boundary. An adversary may exploit vulnerabilities in legitimate agents, inject malicious prompts directly through user-facing channels or indirectly through retrieved documents, tool outputs, or shared artifacts, manipulate intermediate execution context that downstream agents consume, or abuse the legitimate tool-call interface of compromised or coerced agents to steer downstream execution. Crucially, a sophisticated adversary may \emph{launder} malicious intent across multiple agents so that each individual step looks acceptable in isolation~\cite{ma2026safety, he2025red}. The adversary cannot, however, forge agent identities at the runtime layer, compromise the integrity of workflow measurements, or directly tamper with the detector itself; these guarantees are inherited from standard system-security mechanisms and from remote attestation across organizational boundaries.

\textbf{Anomalies in scope.}
Within this trust boundary, we consider two classes of anomalies that both manifest as abnormal workflow execution. \emph{Adversarial workflow violations} arise from intentional manipulations such as prompt injection, tool-description injection, privilege escalation, tool misuse, or coordinated multi-agent compromise. \emph{Intrinsic workflow failures} arise from the system itself, including hallucinated intermediates, flawed planning, ambiguous instructions, coordination breakdowns, runaway retries, and other forms of execution instability. Although their root causes differ, both yield workflows that diverge from the structural and semantic regularities of benign execution; \sysname is designed to detect both under a single decision rule. We further assume that a representative corpus of benign executions is available at training time; the corpus need not be exhaustive, and previously unseen anomalies and zero-day attacks need not appear in any form during training.

\subsection{Problem Definition}
\label{sec:problem}

Consider an agentic AI system of the form described in Sec.~\ref{sec:system-model}. For every user request, the system produces an execution that is recorded, through the workflow construction of Sec.~\ref{sec:system-model}, as a directed attribute graph $\mathcal{G} = (\mathcal{V}, \mathcal{E}, \mathcal{X})$, where $\mathcal{V}$ is the set of observed execution steps (planning actions, tool invocations, inter-agent messages, system events), $\mathcal{E} \subseteq \mathcal{V} \times \mathcal{V}$ is the set of directed dependency edges whose direction encodes both temporal ordering and causal precedence (temporal sequencing, delegation, data flow, and cross-agent communication), and $\mathcal{X}$ collects the node- and edge-level attributes that summarize each step's actor role, action category, invoked tool/endpoint, and execution status, together with edge interaction semantics. Let $\mathcal{G}_{\text{all}}$ denote the space of all such graphs producible by the system. Each executed workflow is modeled as a sample from one of two unknown distributions over $\mathcal{G}_{\text{all}}$: a benign distribution $\mathbb{P}_b$ governing workflows that realize intended task behavior, and an anomalous distribution $\mathbb{P}_a$ governing workflows affected by the adversarial manipulations and intrinsic failures of Sec.~\ref{sec:threat}. At training time, only $\mathbb{P}_b$ is observable, through a finite corpus $\mathcal{D}_b = \{\mathcal{G}_i\}_{i=1}^{N} \stackrel{\mathrm{iid}}{\sim} \mathbb{P}_b$ of measured benign workflows; no labeled anomalies are available.

\textbf{Workflow-level anomaly detection.} Given $\mathcal{D}_b$ and a false-positive budget $\epsilon \in (0,1)$, the goal is to learn a graph-level scoring function $S : \mathcal{G}_{\text{all}} \rightarrow \mathbb{R}_{\ge 0}$ and a threshold $\eta \in \mathbb{R}_{\ge 0}$ that induce the decision rule $\delta_\eta(\mathcal{G}) = \mathbf{1}[\, S(\mathcal{G}) > \eta \,]$, flagging $\mathcal{G}$ as anomalous when its score exceeds $\eta$. We formulate the design problem as
\begin{equation}
\label{eq:problem}
  \min_{S \in \mathcal{S},\, \eta \ge 0} \;
  \mathbb{P}_{\mathcal{G} \sim \mathbb{P}_a}\!\bigl[\, S(\mathcal{G}) \le \eta \,\bigr]
  \quad \text{s.t.} \quad
  \mathbb{P}_{\mathcal{G} \sim \mathbb{P}_b}\!\bigl[\, S(\mathcal{G}) > \eta \,\bigr] \;\le\; \epsilon,
\end{equation}
where $\mathcal{S}$ is a hypothesis class of scoring functions trainable from $\mathcal{D}_b$ without anomaly labels. Because $\mathbb{P}_b$ is accessible only through $\mathcal{D}_b$, the false-positive constraint in~\eqref{eq:problem} is enforced empirically by calibrating $\eta$ as a high quantile of the benign-score distribution $\{S(\mathcal{G}_i)\}_{i=1}^{N}$ at level $q = 1-\epsilon$. The unit of decision is the entire workflow $\mathcal{G}$ rather than any individual step or message, so the problem is intrinsically \emph{graph-level}: $S$ must reflect non-local regularities of benign agentic execution that span multiple steps and multiple agents. The instantiation of $\mathcal{S}$, the construction of $S$, and the calibration of $\eta$ that realize~\eqref{eq:problem} are detailed in Sec.~\ref{sec:methodology}.

\section{The \sysname Framework}
\label{sec:methodology}

\begin{figure*}[t]
  \centering
  \includegraphics[width=0.8\textwidth]{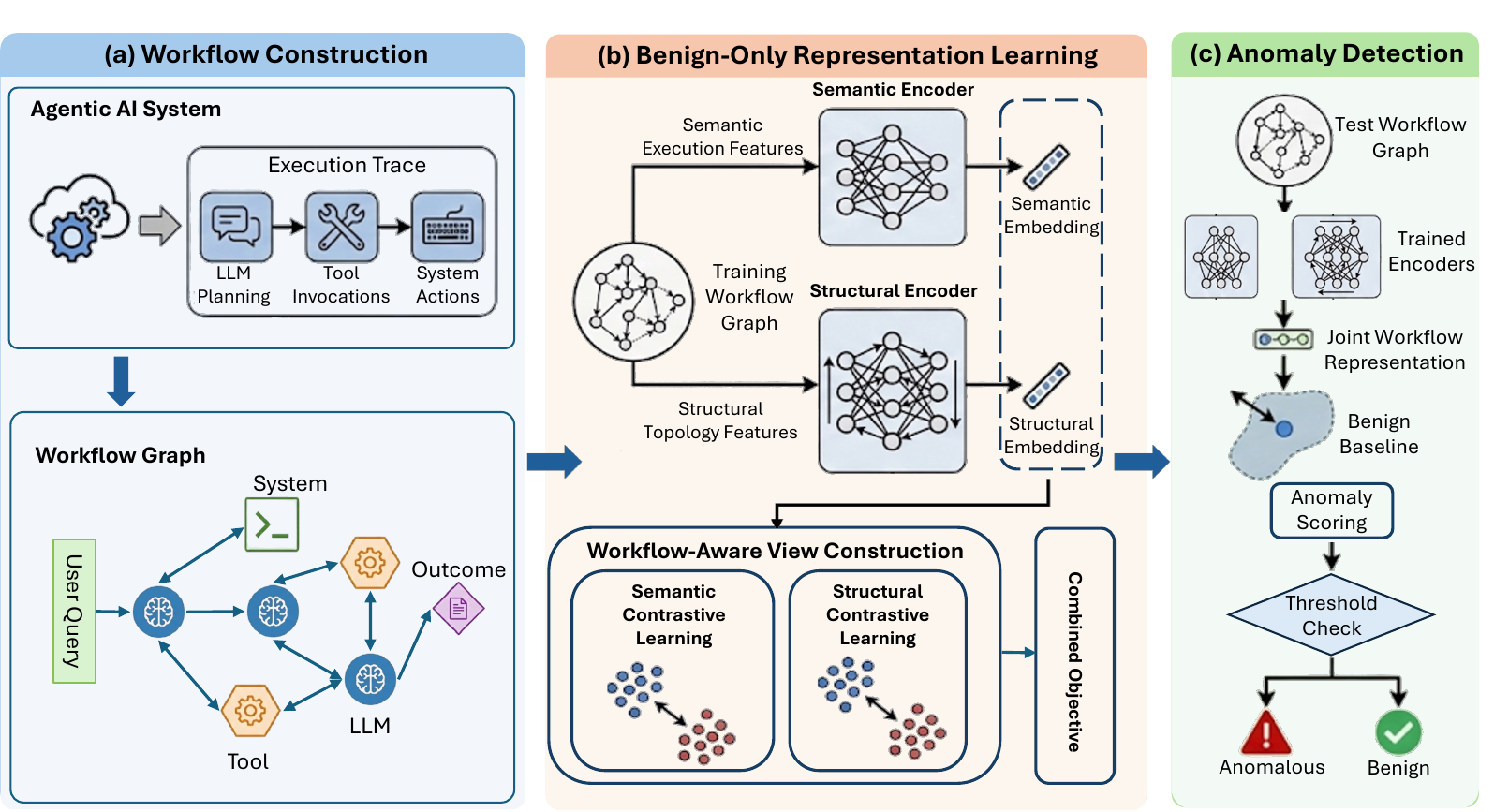}
    \vspace{-0.2cm}
  \caption{\textbf{Overview of \sysname.} \sysname constructs a workflow graph from an observed agentic AI execution, learns benign workflow representations from semantic and structural signals, and detects anomalies by measuring how far a new workflow deviates from learned benign patterns.}
  \Description{Pipeline diagram of Skynet with three stages: workflow graph construction from an observed agentic execution, dual semantic and structural encoding trained on benign workflows, and anomaly scoring by distance to the nearest benign mode anchor.}
  \label{fig:method}
  \vspace{-0.5cm}
\end{figure*}

\subsection{Overview and Design Rationale}
\label{sec:method-overview}

\sysname realizes the workflow-level scoring function $S$, the threshold $\eta$, and the hypothesis class $\mathcal{S}$ sought in problem~\eqref{eq:problem} as an end-to-end pipeline that consumes the workflow graph $\mathcal{G}=(\mathcal{V},\mathcal{E},\mathcal{X})$ defined in Sec.~\ref{sec:problem} and produces a single graph-level decision rule. The pipeline is designed as a \emph{principled} composition rather than a one-off model: every stage is anchored to a property of agentic AI execution, so the detector is benign-only by construction, graph-level by construction, and class-agnostic with respect to whether the underlying anomaly is intrinsic or adversarial.

\emph{Training pipeline} (Figure~\ref{fig:method}) consists of four stages applied to the benign workflow corpus $\mathcal{D}_b=\{\mathcal{G}_i\}_{i=1}^{N}$. \emph{Stage T1: workflow-aware view construction} (Sec.~\ref{sec:method-views}) expands each benign anchor $\mathcal{G}_i$ into an invariant-preserving positive view $\mathcal{G}_i^{+}$ and a set of counterfactual invariant-violating negative views $\mathcal{G}_i^{-}$; both are grown from the structural regularities of benign agentic execution rather than from random graph perturbations, which is what allows the resulting training signal to live near the boundary of benign behavior instead of drifting off into trivially separable noise. \emph{Stage T2: dual-branch encoding} (Sec.~\ref{sec:method-learning}) maps each workflow through a semantic encoder $f_\theta$ that summarizes execution context and a structural encoder $g_\phi$ that summarizes dependency organization, separating \emph{what happens} from \emph{how effects propagate} so that anomalies hiding on either channel remain visible. \emph{Stage T3: workflow-aware contrastive learning} jointly trains $f_\theta$ and $g_\phi$ to pull each anchor toward its invariant-preserving partner and push it away from its counterfactual under a single workflow-level objective derived from agentic workflow invariants. \emph{Stage T4: benign-mode calibration} (Sec.~\ref{sec:method-inference}) consolidates the learned joint embeddings of $\mathcal{D}_b$ into a bank of $K$ benign execution modes $\{c_k\}_{k=1}^{K}$ and sets the threshold $\eta$ to the $q$-quantile of benign training scores with $q=1-\epsilon$, exactly matching the false-positive constraint in~\eqref{eq:problem} without exposure to anomaly labels.

\emph{Inference pipeline} (Figure~\ref{fig:method}, bottom) takes a new workflow $\mathcal{G}$ through the same graph construction and dual-branch encoding, scores it as $S(\mathcal{G}) = \min_{k}\|z(\mathcal{G})-c_k\|_2^2$ in the joint embedding space, and flags it whenever $S(\mathcal{G})>\eta$. The same single rule covers adversarial manipulations, intrinsic failures, and previously unseen anomalies, since every off-manifold workflow surfaces under the same distance, and the procedure runs in two deployment regimes: a per-workflow trailing monitor that scores completed executions, and an incremental per-step monitor that re-scores in-flight executions on every newly observed step. Algorithm~\ref{alg:skynet-main} consolidates the full procedure.

\paragraph{Design rationale.} Three properties of agentic AI execution justify the choices above. \emph{(O1) Dual-channel manifestation.} Agentic anomalies manifest on two distinct channels: many adversarial manipulations preserve the surface vocabulary of agent actions while reshaping dependency flow, and many intrinsic failures preserve the dependency skeleton while corrupting the semantics of intermediate steps; a single representation collapses these channels and misses whichever class hides on the channel it underweights, motivating the dual semantic--structural branches in T2. \emph{(O2) Label-free benign manifold.} Anomaly labels are scarce, inconsistent across deployments, and inherit the blind spots of whichever threat catalog produced them, so the training signal is grown from benign workflows alone, with positives and negatives generated from agentic workflow invariants rather than the random graph augmentations of generic graph contrastive learning, which drift off the benign manifold and yield trivially separable negatives. \emph{(O3) Multimodal benign behavior.} Benign agentic behavior is intrinsically heterogeneous (orchestrator pipelines, ReAct loops, deep-research workflows, and multi-agent debates are all benign yet follow qualitatively different normal regimes), so normality is represented as a set of benign execution modes rather than a single reference, preventing legitimate but atypical workflows from being conflated with anomalies.

\subsection{Workflow-Aware View Construction}
\label{sec:method-views}

The cornerstone of \sysname's benign-only learning is a pair of view-construction operators
$\mathcal{T}^{+}: \mathcal{G}_{\text{all}} \rightarrow \mathcal{G}_{\text{all}}$ and $\mathcal{T}^{-}: \mathcal{G}_{\text{all}} \rightarrow 2^{\mathcal{G}_{\text{all}}}$,
which, for every benign anchor $\mathcal{G}_i \in \mathcal{D}_b$, produce a positive view $\mathcal{G}_i^{+} = \mathcal{T}^{+}(\mathcal{G}_i)$ and a finite set of counterfactual negative views $\mathcal{G}_i^{-} = \mathcal{T}^{-}(\mathcal{G}_i)$. In contrast to the random graph augmentations used in conventional contrastive learning, $\mathcal{T}^{+}$ and $\mathcal{T}^{-}$ are parameterized by \emph{workflow invariants}, that is, properties that benign agentic executions consistently exhibit and that known agentic anomaly classes are known to perturb.

\paragraph{Workflow invariants.}
We identify four such invariants. \emph{(i) Role consistency}: agent responsibilities and delegation patterns remain compatible with their intended functions. \emph{(ii) Provenance and trust consistency}: outputs consumed by later steps originate from expected sources, and benign workflows do not silently elevate the influence of low-trust artifacts on high-stakes downstream actions. \emph{(iii) Dependency and coordination consistency}: prerequisite relations, hand-off structure, and information flow remain coherent, with delegation closing properly and inter-agent coordination converging. \emph{(iv) Stability}: executions remain bounded and avoid uncontrolled retries, loops, or oscillatory propagation. Each invariant is empirically associated with one or more known anomaly families: prompt injection commonly violates role and provenance consistency~\cite{debenedetti2024agentdojo, zhan2024injecagent}; coordination breakdowns and intrinsic planning failures violate dependency consistency~\cite{zhou2025guardian}; runaway retries and progress-misjudgment failures violate stability~\cite{zhu2025llmagenterror}. Anchoring view construction on these invariants is what makes \sysname workflow-aware rather than a generic graph learner, and what gives the resulting training signal a semantic interpretation grounded in the agentic-AI threat surface.

\paragraph{Invariant-preserving views (positives).}
The positive operator $\mathcal{T}^{+}$ maps $\mathcal{G}_i$ to a view $\mathcal{G}_i^{+}$ that respects all four invariants while exposing the natural variability of benign agentic execution. We instantiate $\mathcal{T}^{+}$ as a stochastic composition of three families of transformations: (i) \emph{observation-preserving} rewrites retain the dependency structure while masking or perturbing non-essential observation content (e.g., paraphrasing free-form notes, redacting irrelevant tool fields); (ii) \emph{execution-equivalent} rewrites apply local edits that preserve task semantics, such as reordering independent operations or substituting semantically equivalent tool/action attributes; and (iii) \emph{prefix or partial} rewrites expose incomplete yet still valid sub-workflows, reflecting the partial observability faced by an in-flight monitor. By construction $\mathcal{G}_i^{+}$ remains in $\mathrm{supp}(\mathbb{P}_b)$ and is therefore a \emph{semantically faithful} positive: pulling $\mathcal{G}_i$ and $\mathcal{G}_i^{+}$ together in representation space teaches the encoders to be invariant to the natural variability of benign agentic execution rather than to arbitrary graph noise.

\paragraph{Counterfactual invariant-violating views (negatives).}
The negative operator $\mathcal{T}^{-}$ is the design move that distinguishes \sysname's contrastive scheme from generic graph contrastive learning. For each benign anchor $\mathcal{G}_i$ and invariant $I \in \{\textsc{Role}, \textsc{Prov}, \textsc{Dep}, \textsc{Stab}\}$, an invariant-specific operator $\mathcal{T}^{-}_{I}$ produces a counterfactual $\mathcal{G}_i^{-,I}$ that minimally violates $I$ while leaving the remaining three invariants intact. Concretely, role counterfactuals reassign or relabel an agent's action so that a delegation crosses a benign role boundary; provenance counterfactuals reroute one downstream consumption to draw from a non-canonical source; dependency counterfactuals delete or insert a single hand-off edge so that a sub-workflow no longer closes properly; stability counterfactuals extend a retry or loop tail beyond benign envelopes. Each $\mathcal{G}_i^{-,I}$ obeys three principles by design: \emph{minimal modification} (smallest perturbation sufficient to violate the targeted invariant), \emph{single-invariant violation} (the training signal isolates one axis of agentic anomaly), and \emph{local plausibility} (the perturbed workflow could plausibly arise in real agentic execution). The full negative set $\mathcal{G}_i^{-} \;=\; \bigcup_{I}\, \bigl\{\mathcal{G}_i^{-,I}\bigr\}$.
Therefore, it carves out the \emph{boundary} of $\mathrm{supp}(\mathbb{P}_b)$ along agentic-meaningful directions, in contrast to random edge or attribute perturbations whose negatives drift far outside $\mathrm{supp}(\mathbb{P}_b)$ and are trivially separable. The ablation in Sec.~\ref{sec:eval-ablation} confirms that replacing $\mathcal{T}^{-}$ with random graph augmentation collapses recall on near-manifold attacks and intrinsic failures even when the rest of the pipeline is held fixed.

\subsection{Dual-Branch Workflow Representation Learning}
\label{sec:method-learning}

\paragraph{Two-branch encoder.}
Given a workflow graph $\mathcal{G}$, \sysname computes two embeddings,
$h(\mathcal{G}) = f_{\theta}(\mathcal{G}) \in \mathbb{R}^{d_h}$ and $t(\mathcal{G}) = g_{\phi}(\mathcal{G}) \in \mathbb{R}^{d_t}$,
where the \emph{semantic encoder} $f_\theta$ is a message-passing graph encoder that consumes node and edge attributes (actor role, action category, tool/endpoint, status, edge type), and the \emph{structural encoder} $g_\phi$ is a message-passing graph encoder that consumes the structural descriptors of each node (degree statistics, distances to workflow boundaries, ancestor/descendant relations, and influence-concentration scores). The two branches do not share parameters; this decoupling is intentional and reflects observation~(O1): agentic anomalies frequently perturb only one of the two channels, and parameter sharing would force a single representation to compromise between them.

\paragraph{Per-anchor contrast pool.}
Let $\mathcal{B} \subseteq \mathcal{D}_b$ be a mini-batch of size $B$. For each anchor $\mathcal{G}_i \in \mathcal{B}$, we instantiate $\mathcal{G}_i^{+} = \mathcal{T}^{+}(\mathcal{G}_i)$ and $\mathcal{G}_i^{-} = \mathcal{T}^{-}(\mathcal{G}_i)$ on the fly and define the per-anchor contrast pool $\mathcal{Q}_i = \{\mathcal{G}_j^{+} : j \in \mathcal{B},\, j \neq i\} \cup \mathcal{G}_i^{-}$, which combines in-batch benign negatives with same-anchor counterfactual negatives. The two sources are complementary: in-batch benign negatives spread different benign workflows across the embedding space and prevent representation collapse, while same-anchor counterfactual negatives sharpen the decision surface along the agentic-invariant directions that downstream anomalies actually perturb.

\paragraph{Dual contrastive objective.}
Let $\mathrm{sim}(u,v) = u^{\top} v / (\|u\|\,\|v\|)$ denote cosine similarity and $\tau>0$ a temperature. The semantic and structural branches are trained under analogous standard contrastive objectives,
\begin{align}
\label{eq:loss-sem}
L_{\text{sem}}(\theta) &\;=\; -\,\frac{1}{B}\sum_{i=1}^{B} \log
\frac{\exp\bigl(\mathrm{sim}(h_i,\,h_i^{+})/\tau\bigr)}
     {\displaystyle\sum_{\mathcal{G}\in\{\mathcal{G}_i^{+}\}\cup\mathcal{Q}_i}\!\!\exp\bigl(\mathrm{sim}(h_i,\,h(\mathcal{G}))/\tau\bigr)}, \\[2pt]
\label{eq:loss-str}
L_{\text{str}}(\phi) &\;=\; -\,\frac{1}{B}\sum_{i=1}^{B} \log
\frac{\exp\bigl(\mathrm{sim}(t_i,\,t_i^{+})/\tau\bigr)}
     {\displaystyle\sum_{\mathcal{G}\in\{\mathcal{G}_i^{+}\}\cup\mathcal{Q}_i}\!\!\exp\bigl(\mathrm{sim}(t_i,\,t(\mathcal{G}))/\tau\bigr)},
\end{align}
where $h_i = h(\mathcal{G}_i)$, $h_i^{+} = h(\mathcal{G}_i^{+})$, and $t_i, t_i^{+}$ are defined analogously. The two branches are combined into the workflow-level training objective
\begin{equation}
\label{eq:loss-total}
L(\theta,\phi) \;=\; \alpha\, L_{\text{sem}}(\theta) \;+\; \beta\, L_{\text{str}}(\phi),
\qquad \alpha,\beta>0,
\end{equation}
which is minimized by stochastic gradient descent over the two encoders' parameters. Intuitively, $L_{\text{sem}}$ teaches $f_\theta$ that two workflows with the same task semantics but different surface presentation should embed close, and that a workflow whose role, provenance, dependency, or stability invariant has been minimally violated should embed far; $L_{\text{str}}$ enforces the analogous property in the structural channel. Concretely, the gradient of $L_{\text{sem}}$ with respect to $h_i$ is a softmax-weighted sum that pulls $h_i$ toward $h_i^{+}$ and pushes it away from each $h(\mathcal{G})$ for $\mathcal{G}\in\mathcal{Q}_i$, with the strongest repulsion applied to whichever counterfactual currently sits closest to the anchor in representation space; the same dynamics hold for $L_{\text{str}}$ in the structural channel. Through the standard mutual-information lower-bound interpretation of contrastive learning, minimizing~\eqref{eq:loss-sem}--\eqref{eq:loss-str} maximizes a tractable lower bound on $I\bigl(h(\mathcal{G});\,h(\mathcal{G}^{+})\bigr)$ and $I\bigl(t(\mathcal{G});\,t(\mathcal{G}^{+})\bigr)$, so the encoders retain exactly the information that is invariant under the benign view operator $\mathcal{T}^{+}$ and discard the information that the counterfactual operator $\mathcal{T}^{-}$ shows is \emph{not} invariant under benign behavior. The novelty of~\eqref{eq:loss-total} therefore lies not in the contrastive form itself but in \emph{what it contrasts}: invariant-preserving and invariant-violating views, applied along a dual semantic--structural decomposition that mirrors the two channels through which agentic anomalies manifest.

\paragraph{Hypothesis class.}
The hypothesis class $\mathcal{S}$ from problem~\eqref{eq:problem} is therefore realized as $\mathcal{S} = \{\, \mathcal{G} \mapsto \min_{k} \| z_{\theta,\phi}(\mathcal{G}) - c_k \|_2^{2} \mid (\theta,\phi)\in\Theta,\;\{c_k\}_{k=1}^{K}\subset\mathbb{R}^{d_h+d_t} \,\}$, with $z_{\theta,\phi}(\mathcal{G}) = [\, h(\mathcal{G})\,;\,\sqrt{\lambda}\,t(\mathcal{G})\,]$. The encoder parameters $(\theta,\phi)$ are fit by minimizing~\eqref{eq:loss-total}; the mode anchors $\{c_k\}$ and threshold $\eta$ are fit subsequently on $\mathcal{D}_b$ alone, as described next.

\subsection{Workflow-Mode Anomaly Scoring}
\label{sec:method-inference}

After training, \sysname forms the joint workflow representation $z(\mathcal{G}) = [\, h(\mathcal{G})\,;\,\sqrt{\lambda}\,t(\mathcal{G})\,] \in \mathbb{R}^{d_h+d_t}$, with $\lambda>0$ rescaling the structural branch. To respect the heterogeneity of benign agentic execution (O3), normality is modeled as a set of $K$ \emph{benign execution modes} $\{c_k\}_{k=1}^{K}$ obtained by $k$-means on $\{z(\mathcal{G}_i)\}_{i=1}^{N}$, where each mode corresponds to a recurrent pattern of nominal behavior such as a coordination style, a tool-usage regime, or a task-decomposition structure. The graph-level anomaly score $S$ in problem~\eqref{eq:problem} is then instantiated as the squared distance to the nearest benign mode,
\[
  S(\mathcal{G}) \;=\; \min_{1\le k\le K}\, \bigl\| z(\mathcal{G}) - c_k \bigr\|_{2}^{2},
\]
and the threshold $\eta$ is calibrated as the empirical $q$-quantile of $\{S(\mathcal{G}_i)\}_{i=1}^{N}$ with $q = 1-\epsilon$, exactly enforcing the false-positive constraint of~\eqref{eq:problem} on $\mathcal{D}_b$. A test workflow is flagged when $S(\mathcal{G}) > \eta$. Because $S$ measures distance to the benign manifold rather than similarity to any specific anomaly family, the same single decision rule covers adversarial manipulations, intrinsic failures, and previously unseen anomalies: any workflow that fits none of the benign modes surfaces as off-manifold geometry under $S$. The end-to-end procedure is summarized in Algorithm~\ref{alg:skynet-main}.

\begin{algorithm}[t]
\caption{\sysname: benign-only workflow-level anomaly detection.}
\label{alg:skynet-main}
\begin{algorithmic}[1]
\REQUIRE Benign workflow-graph corpus $\mathcal{D}_{b}=\{\mathcal{G}_i\}_{i=1}^{N}$; semantic encoder $f_{\theta}$; structural encoder $g_{\phi}$; loss weights $\alpha,\beta$; structural scale $\lambda$; temperature $\tau$; benign-mode count $K$; calibration quantile $q$; training epochs $E$; batch size $B$.
\ENSURE Trained encoders $(f_{\theta}, g_{\phi})$; benign-mode bank $\{c_k\}_{k=1}^{K}$; detection threshold $\eta$; anomaly scorer $S(\cdot)$.
\STATE Initialize $\theta, \phi$.
\FOR{$\text{epoch}=1$ \textbf{to} $E$}
  \STATE Sample a mini-batch $\mathcal{B}\subset\mathcal{D}_{b}$ of size $B$.
  \STATE For each $\mathcal{G}_i\in\mathcal{B}$, generate an invariant-preserving view $\mathcal{G}_i^{+}$ and a set of counterfactual views $\mathcal{G}_i^{-}$ that each minimally violate one role, provenance, dependency, or stability invariant.
  \STATE Form the per-anchor contrast pool $\mathcal{Q}_i \leftarrow \{\mathcal{G}_j^{+}\}_{j\neq i}\cup \mathcal{G}_i^{-}$.
  \STATE Compute $h_i, h_i^{+}, t_i, t_i^{+}$ for $\mathcal{G}_i, \mathcal{G}_i^{+}$ and $h(\mathcal{G}),t(\mathcal{G})$ for all $\mathcal{G}\in\mathcal{Q}_i$.
  \STATE $L_{\mathrm{sem}}\leftarrow -\tfrac{1}{B}\sum_i \log\frac{\exp(\mathrm{sim}(h_i,h_i^{+})/\tau)}{\sum_{\mathcal{G}\in\{\mathcal{G}_i^{+}\}\cup\mathcal{Q}_i}\exp(\mathrm{sim}(h_i,h(\mathcal{G}))/\tau)}$.
  \STATE $L_{\mathrm{str}}\leftarrow -\tfrac{1}{B}\sum_i \log\frac{\exp(\mathrm{sim}(t_i,t_i^{+})/\tau)}{\sum_{\mathcal{G}\in\{\mathcal{G}_i^{+}\}\cup\mathcal{Q}_i}\exp(\mathrm{sim}(t_i,t(\mathcal{G}))/\tau)}$.
  \STATE $L\leftarrow \alpha L_{\mathrm{sem}}+\beta L_{\mathrm{str}}$; update $\theta, \phi$ by one optimizer step on $L$.
\ENDFOR
\STATE Compute $\{z(\mathcal{G}_i)=[h(\mathcal{G}_i);\sqrt{\lambda}\,t(\mathcal{G}_i)]\}_{i=1}^{N}$ and fit $K$ benign-mode anchors $\{c_k\}_{k=1}^{K}$ by $k$-means in the joint embedding space.
\STATE Define $S(\mathcal{G}) \leftarrow \min_k \|z(\mathcal{G})-c_k\|_2^{2}$; set $\eta$ to the $q$-quantile of $\{S(\mathcal{G}_i)\}_{i=1}^{N}$.
\STATE \textbf{At inference}, given a (possibly in-flight) test workflow prefix $\mathcal{G}$:
\STATE \quad compute $z(\mathcal{G})=[f_\theta(\mathcal{G});\sqrt{\lambda}\,g_\phi(\mathcal{G})]$ and $s\leftarrow S(\mathcal{G})$.
\STATE \quad \textbf{if} $s>\eta$ \textbf{then} flag $\mathcal{G}$ as anomalous \textbf{else} accept as benign.
\STATE \quad The same procedure is applied either once on the completed workflow (trailing per-workflow mode) or after each newly observed step on the current prefix (incremental per-step mode).
\STATE \textbf{return} $(f_\theta, g_\phi, \{c_k\}, \eta, S)$.
\end{algorithmic}
\end{algorithm}

\section{Evaluation}
\label{sec:evaluation}

We evaluate \sysname as an always-on workflow-level anomaly detector around three questions.
\textbf{RQ1.} Does it detect workflow-level anomalies across both adversarial manipulations and intrinsic failures while sustaining a deployment-relevant low-FPR operating point (Sec.~\ref{sec:eval-effectiveness})?
\textbf{RQ2.} Does each workflow-aware learning component deliver the operating-point gain it is designed for (Sec.~\ref{sec:eval-ablation})?
\textbf{RQ3.} Is its runtime overhead compatible with always-on monitoring (Sec.~\ref{sec:eval-efficiency})?

\subsection{Evaluation Setup}
\label{sec:eval-setup}

\textbf{Benchmarks.}
For adversarial evaluation we use Agent-SafetyBench \cite{zhang2024agent} ($111$ trajectories from safety red-teaming) and ATBench~\cite{li2026atbench} ($497$ trajectories across eight risk sources and fourteen failure modes); for intrinsic failures we use AgentErrorBench~\cite{zhu2025llmagenterror} ($200$ non-adversarial trajectories across AlfWorld, GAIA, and WebShop). A \emph{Combined} scenario aggregates all $808$ anomalies with the benign test set to measure cross-source robustness when the anomaly distribution broadens. Benign trajectories come from \textsc{ETO}~\cite{song2024eto} on AlfWorld~\cite{shridharalfworld}, SciWorld~\cite{wang2022sciworld}, and WebShop~\cite{yao2022webshop}, converted to workflow graphs (Sec.~\ref{sec:methodology}) and partitioned into $4{,}494$/$562$/$562$ train/validation/test splits; anomalies are never seen during training. The three anomaly benchmarks are public datasets independently released by different groups; using them together avoids co-design bias between training and evaluation.

\textbf{Baselines.}
We compare against six detectors covering the current design space: agent-centric textual monitoring (\textsc{GuardAgent} \cite{xiang2025guardagent}); provenance-level defenses (\textsc{TraceAegis} \cite{liu2025traceaegis}, \textsc{Guardian} \cite{zhou2025guardian}); and graph anomaly detectors (\textsc{F-GAE} \cite{feng2022fgae}, \textsc{CoLA} \cite{liu2022cola}, \textsc{TopoGCL}\cite{chen2024topogcl}). All baselines share the same graph inputs and benign training split, each tuning its own threshold via a balance-oriented validation rule. To prevent trivial detection via surface features, graph baselines are stripped of text- and length-based shortcuts: raw text content is removed from node attributes and only categorical/numerical workflow features (role, action type, tool, status, structural descriptors) are retained, and per-graph node/edge counts are normalized so that graph-size cues alone cannot drive the score.

\textbf{Implementation and metrics.}
We set structural scale $\lambda=1.2$ and combine the semantic and structural objectives with $\alpha=0.6$, $\beta=0.4$. Normality is modeled by a benign-only $k$-means bank with $K=150$ modes, and the threshold is calibrated on benign training scores at a quantile that targets a sub-$1\%$ benign false-positive budget, yielding a single fixed operating point that materializes to $\approx 0.89\%$ FPR on the held-out benign test split. We report accuracy, precision, recall, F1, and FPR; FPR is the deployment-critical metric, since any always-on monitor firing above a few percent FPR is unusable in production. ROC curves provide a threshold-independent view that controls for the asymmetry between \sysname's fixed-quantile calibration and the per-method balance-oriented thresholds used by the baselines.

\subsection{Detection Effectiveness (RQ1)}
\label{sec:eval-effectiveness}

\begin{figure}[t]
    \centering
    \includegraphics[width=0.9\linewidth]{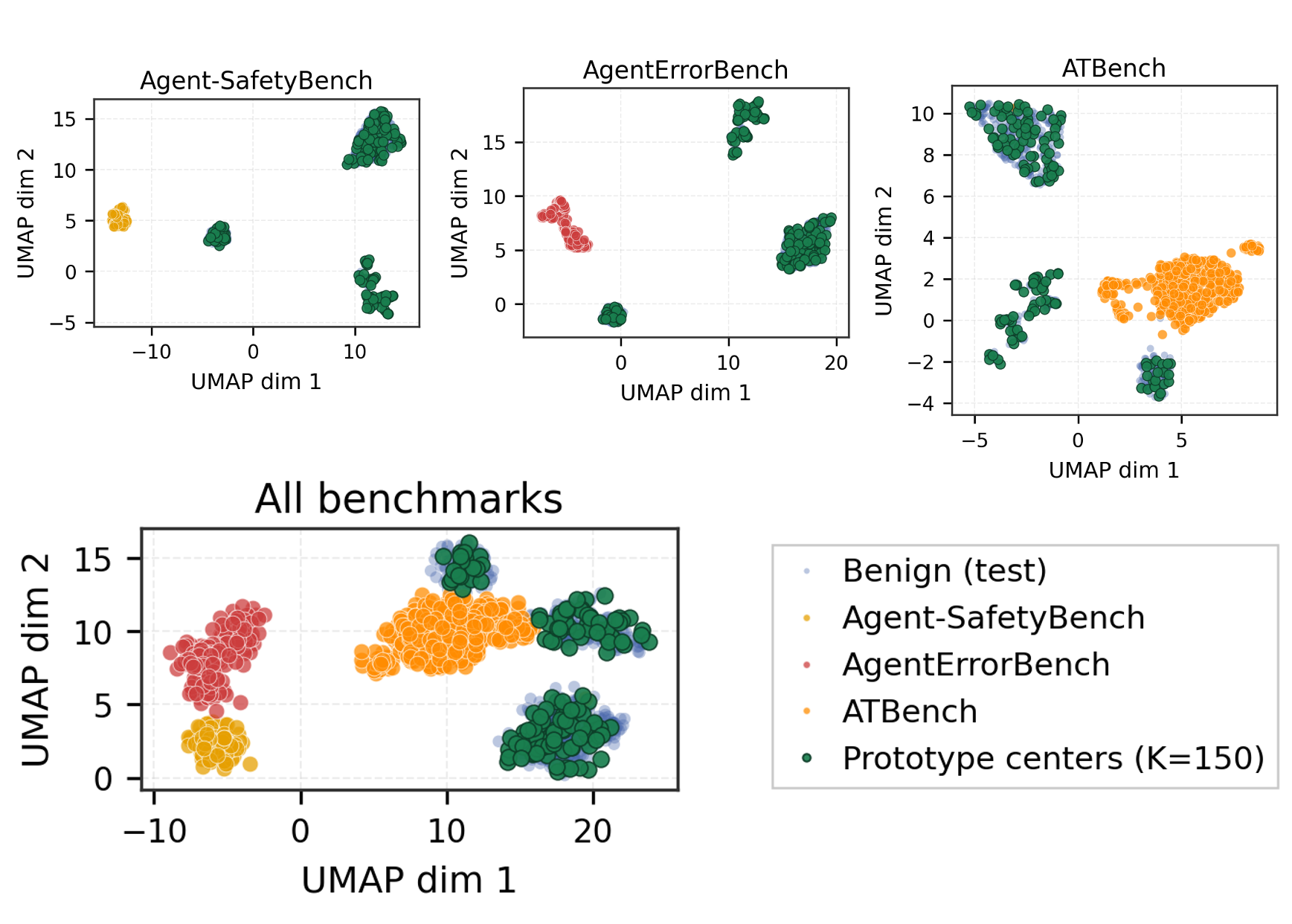}
    \vspace{-0.5cm}
    \caption{UMAP projection of the learned joint workflow embedding $z(\mathcal{G})$. Top-row panels overlay anomalies from each benchmark on the held-out benign test set; the bottom panel shows all three benchmarks together with the $K{=}150$ benign mode anchors. Benign workflows form a coherent manifold around the anchors; anomalies from each source consistently lie off-manifold.}
    \vspace{-0.5cm}
    
    \Description{UMAP scatter plots of workflow embeddings. Benign workflows cluster tightly around the benign mode anchors, while anomalous workflows from each benchmark scatter away from that cluster.}
    \label{fig:umap}
\end{figure}

\textbf{Embedding-space geometry.}
Figure~\ref{fig:umap} qualitatively illustrates how \sysname reshapes the workflow space: benign workflows concentrate in a coherent manifold around the mode anchors, while anomalies from every benchmark lie off-manifold regardless of which threat class produced them. This is the geometry the multi-mode detector exploits. UMAP is nonlinear and stochastic, so the 2D view is illustrative; all detection scores below are computed in the original joint embedding.

\begin{table*}[t]
\centering
\small
\setlength{\tabcolsep}{3pt}
\resizebox{\textwidth}{!}{%
\begin{tabular}{l ccccc ccccc ccccc ccccc}
\toprule
& \multicolumn{5}{c}{Agent-SafetyBench} & \multicolumn{5}{c}{AgentErrorBench} & \multicolumn{5}{c}{ATBench} & \multicolumn{5}{c}{Combined} \\
\cmidrule(lr){2-6} \cmidrule(lr){7-11} \cmidrule(lr){12-16} \cmidrule(lr){17-21}
Method & Acc & Prec & Rec & F1 & FPR & Acc & Prec & Rec & F1 & FPR & Acc & Prec & Rec & F1 & FPR & Acc & Prec & Rec & F1 & FPR \\
\midrule
\textbf{\sysname}     & 99.26 & 95.69 & \textbf{100.00} & 97.80 & 0.89 & 99.34 & 97.56 & \textbf{100.00} & 98.77 & 0.89 & \textbf{91.41} & 98.80 & 82.70 & \textbf{90.03} & 0.89 & \textbf{93.36} & 99.31 & \textbf{89.36} & \textbf{94.07} & 0.89 \\
\textsc{F-GAE}        & 97.86 & 90.91 & 99.10 & 94.83 & 2.14 & 98.05 & 93.02 & 98.50 & 95.68 & 2.14 & 88.94 & 90.12 & \textbf{84.31} & 87.12 & 2.14 & 90.87 & 93.44 & 86.92 & 90.06 & 2.14 \\
\textsc{TraceAegis}   & 96.62 & 86.89 & 96.40 & 91.40 & 3.38 & 96.83 & 89.19 & 94.50 & 91.77 & 3.38 & 67.43 & 61.54 & 34.61 & 44.27 & 3.38 & 75.92 & 78.48 & 58.42 & 66.98 & 3.38 \\
\textsc{Guardian}     & \textbf{99.85} & \textbf{99.11} & \textbf{100.00} & \textbf{99.55} & \textbf{0.18} & \textbf{99.87} & \textbf{99.50} & \textbf{100.00} & \textbf{99.75} & \textbf{0.18} & 61.57 & \textbf{98.91} & 18.31 & 30.90 & \textbf{0.18} & 70.29 & \textbf{99.75} & 49.75 & 66.39 & \textbf{0.18} \\
\textsc{TopoGCL}      & 95.69 & 80.15 & 98.20 & 88.26 & 4.80 & 92.78 & 86.43 & 86.00 & 86.22 & 4.80 & 56.56 & 70.33 & 12.88 & 21.77 & 4.80 & 64.23 & 92.74 & 42.70 & 58.47 & 4.80 \\
\textsc{CoLA}         & 95.10 & 77.08 & \textbf{100.00} & 87.06 & 5.87 & 71.26 & 29.79 & 7.00 & 11.34 & 5.87 & 73.56 & 88.34 & 50.30 & 64.10 & 5.87 & 65.99 & 91.91 & 46.41 & 61.68 & 5.87 \\
\textsc{GuardAgent}   & 84.10 & 64.29 & 8.11 & 14.40 & 0.89 & 73.10 & 0.00 & 0.00 & 0.00 & 0.89 & 60.25 & 94.19 & 16.30 & 27.79 & 0.89 & 47.23 & 94.74 & 11.14 & 19.93 & 0.89 \\
\bottomrule
\end{tabular}%
}
\caption{Detection performance across the four evaluation scenarios (all values in \%).}
\vspace{-0.7cm}
\label{tab:baselines}
\end{table*}

\begin{figure*}[t]
    \centering
    \includegraphics[width=0.8\linewidth]{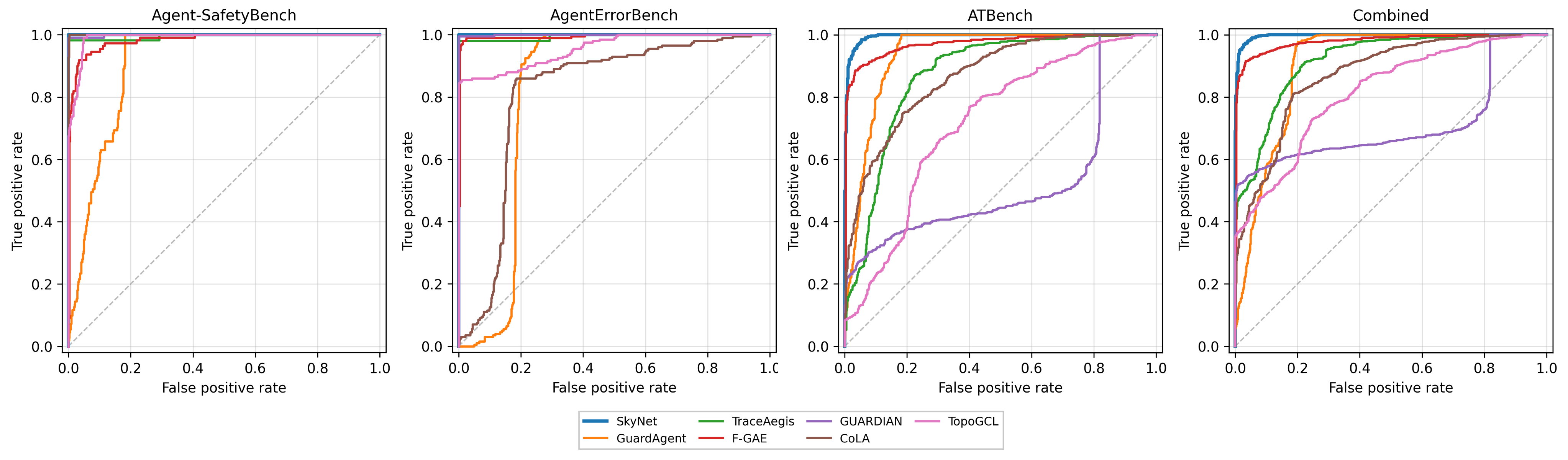}
    \caption{ROC curves of \sysname and the six baselines across the four evaluation scenarios. \sysname dominates the deployment-relevant low-FPR region across all four scenarios; baselines that are near-perfect on Agent-SafetyBench and AgentErrorBench degrade sharply on ATBench and Combined.}
    \vspace{-0.5cm}
    \Description{Four ROC plots, one per evaluation scenario, comparing Skynet against six baselines. The Skynet curve rises fastest in the low false-positive-rate region of every panel.}
    \label{fig:roc}
\end{figure*}

\textbf{Cross-scenario detection.}
Table~\ref{tab:baselines} and Figure~\ref{fig:roc} show that \sysname is the only detector that maintains high recall with sub-$1\%$ FPR across all four scenarios at one fixed operating point, as required for online monitoring. It achieves $100\%$ recall at $0.89\%$ FPR on Agent-SafetyBench and AgentErrorBench, delivers the best recall--FPR trade-off on ATBench and Combined, and attains the highest F1 on both broader scenarios. The Combined setting most clearly distinguishes workflow-level monitoring from prior approaches. On the focused benchmarks, several baselines also achieve near-saturated recall below $1\%$ FPR (e.g., \textsc{Guardian} matches \sysname's recall at lower FPR), as coarse trace or coordination shifts are easy for graph monitors to detect. Under broader threat distributions, however, they degrade sharply: \textsc{Guardian}'s recall falls to the tens of percent on ATBench and roughly halves on Combined, while \textsc{TraceAegis} loses most recall on ATBench, indicating poor transfer of coordination- or trace-specific signals. In contrast, \sysname detects off-manifold deviations from benign workflow regularities, allowing it to transfer across heterogeneous adversarial and intrinsic anomalies while preserving both high recall and the sub-$1\%$ FPR budget, also reflected in the highest mean F1. The ROC curves confirm this threshold-independently: text-only monitoring is insufficient at workflow scale; trace- and provenance-level defenses perform well on focused sets but degrade under distribution shift; and generic graph anomaly detectors require $2$--$6\%$ FPR for competitive recall, making them unsuitable for always-on monitoring.

\begin{table}[t]
\centering
\small
\setlength{\tabcolsep}{3pt}
\begin{tabular}{l r cccc c}
\toprule
\textbf{Subclass} & $\boldsymbol{n}$ & \textbf{Acc} & \textbf{Prec} & \textbf{Rec} & \textbf{F1} & \textbf{FPR} \\
\midrule
\multicolumn{7}{l}{\textbf{AgentErrorBench (failure type)}} \\
inefficient\_plan            & 46 & 99.18 & 90.20 & 100.00 & 94.85 & 0.89 \\
over\_simplification         & 20 & 99.14 & 80.00 & 100.00 & 88.89 & 0.89 \\
progress\_misjudge           & 20 & 99.14 & 80.00 & 100.00 & 88.89 & 0.89 \\
constraint\_ignorance        & 13 & 99.13 & 72.22 & 100.00 & 83.87 & 0.89 \\
hallucination                & 13 & 99.13 & 72.22 & 100.00 & 83.87 & 0.89 \\
outcome\_misinterpretation   & 12 & 99.13 & 70.59 & 100.00 & 82.76 & 0.89 \\
impossible\_action           & 10 & 99.13 & 66.67 & 100.00 & 80.00 & 0.89 \\
step\_limit                  &  8 & 99.12 & 61.54 & 100.00 & 76.19 & 0.89 \\
environment\_error           &  7 & 99.12 & 58.33 & 100.00 & 73.68 & 0.89 \\
misalignment                 &  7 & 99.12 & 58.33 & 100.00 & 73.68 & 0.89 \\
causal\_misattribution       &  5 & 99.12 & 50.00 & 100.00 & 66.67 & 0.89 \\
parameter\_error             &  4 & 99.12 & 44.44 & 100.00 & 61.54 & 0.89 \\
tool\_execution\_error       &  4 & 99.12 & 44.44 & 100.00 & 61.54 & 0.89 \\
memory\_retrieval\_failure   &  2 & 99.11 & 28.57 & 100.00 & 44.44 & 0.89 \\
invalid\_action              &  1 & 99.11 & 16.67 & 100.00 & 28.57 & 0.89 \\
llm\_limit                   &  1 & 99.11 & 16.67 & 100.00 & 28.57 & 0.89 \\
unknown                      & 27 & 99.15 & 84.38 & 100.00 & 91.53 & 0.89 \\
\midrule
\multicolumn{7}{l}{\textbf{ATBench (risk source)}} \\
direct\_prompt\_injection      &  39 & 98.84 & 88.10 &  94.87 & 91.36 & 0.89 \\
indirect\_prompt\_injection    &  77 & 96.09 & 91.94 &  74.03 & 82.01 & 0.89 \\
tool\_description\_injection   &  52 & 98.37 & 90.38 &  90.38 & 90.38 & 0.89 \\
jailbreak                      &  61 & 97.75 & 91.23 &  85.25 & 88.14 & 0.89 \\
unreliable\_or\_misinformation &  64 & 97.28 & 91.23 &  81.25 & 85.95 & 0.89 \\
corrupted\_tool\_feedback      &  44 & 98.35 & 88.64 &  88.64 & 88.64 & 0.89 \\
malicious\_tool\_execution     &  25 & 99.15 & 83.33 & 100.00 & 90.91 & 0.89 \\
inherent\_agent\_failures      & 135 & 94.55 & 95.33 &  75.56 & 84.30 & 0.89 \\
\bottomrule
\end{tabular}
\caption{Fine-grained detection performance of \sysname on AgentErrorBench failure types and ATBench risk sources (all values in \%).}
\vspace{-0.7cm}
\label{tab:subclass}
\end{table}

\textbf{Generality across agentic anomaly regimes.}
Table~\ref{tab:subclass} breaks down failure types and adversarial behaviors under the same fixed detector. On AgentErrorBench, \sysname achieves full recall across all $17$ intrinsic failure types, including subtle errors such as hallucination, constraint ignorance, and causal misattribution, indicating a broadly discriminative boundary over failures unseen during training. Lower precision on rare categories is largely a sample-size effect: at the fixed $0.89\%$ FPR, the $\sim$5 benign false positives among $562$ samples dominate precision when only a few positives are present. On ATBench, recall remains high for overt attacks (e.g., \emph{malicious\_tool\_execution} and \emph{direct\_prompt\_injection}) but drops for \emph{inherent\_agent\_failures} and \emph{indirect\_prompt\_injection}, whose executions most closely resemble benign workflows. These cases mark the limit of workflow-level off-manifold detection and motivate complementary content-aware checks.

\subsection{Workflow-Aware Learning Scheme: Component Contributions (RQ2)}
\label{sec:eval-ablation}

\begin{table}[t]
\centering
\small
\setlength{\tabcolsep}{4pt}
\begin{tabular}{l ccccc}
\toprule
\textbf{Variant} & \textbf{Acc} & \textbf{Prec} & \textbf{Rec} & \textbf{F1} & \textbf{FPR} \\
\midrule
Full \sysname                & \textbf{96.64} & \textbf{99.48} & \textbf{94.80} & \textbf{97.08} & \textbf{0.71} \\
\;w/o structural branch      & 94.09 & 99.32 & 90.59 & 94.76 & 0.89 \\
\;w/o counterfactual views   & 94.09 & 99.19 & 90.72 & 94.76 & 1.07 \\
\;w/o workflow-aware aug.    & 95.84 & 99.30 & 93.54 & 96.33 & 0.98 \\
\;single-mode inference      & 79.42 & 99.25 & 65.59 & 78.99 & \textbf{0.71} \\
\bottomrule
\end{tabular}
\caption{Component-level ablation on the Combined scenario, with one component disabled at a time (all values in \%).}
\vspace{-0.7cm}
\label{tab:ablation}
\end{table}

Table~\ref{tab:ablation} quantifies each component's contribution on Combined using the same data as Table~\ref{tab:baselines}, but from a separate ablation run. We prioritize recall and FPR, using F1 only as a secondary check, since even modest recall loss or a few-percent FPR increase is costly for always-on monitoring. \emph{Structural branch.} Removing it reduces recall by several points and raises FPR to $\sim$0.9\%, with a corresponding F1 drop, showing that dependency-sensitive structure adds information beyond semantics; many agentic failures arise from how effects propagate through a workflow rather than from local step content. \emph{Counterfactual invariant-violating views.} Removing them increases FPR by roughly $1.5\times$, exceeding the $1\%$ budget, while only modestly reducing recall, indicating that they primarily tighten the benign boundary against near-manifold deviations. \emph{Multi-mode benign scoring.} Collapsing the $K{=}150$ benign modes into a single reference is the most damaging ablation: recall drops by roughly one third while FPR remains similar, highlighting both the multimodality of benign workflows and the limitation of single-reference one-class scoring. \emph{Workflow-aware view construction.} Removing it lowers recall and raises FPR by about $1.4\times$ beyond the $1\%$ budget, confirming that physically realizable workflow views improve both detection and calibration. Overall, the components are complementary: structural encoding and counterfactual views shape off-manifold geometry, multi-mode scoring captures benign diversity, and workflow-aware views preserve the low-FPR regime required for always-on monitoring.

\subsection{Always-On Monitoring Overhead (RQ3)}
\label{sec:eval-efficiency}

\begin{table}[t]
\centering
\small
\setlength{\tabcolsep}{4pt}
\begin{tabular}{l rrr}
\toprule
\textbf{Component} & \textbf{Total (s)} & \textbf{Mean (s)} & \textbf{P95 (s)} \\
\midrule
\multicolumn{4}{l}{\emph{Per-workflow component cost (offline)}} \\
Graph construction                  & 0.2249 & 0.0225 & 0.0882 \\
Encoder forward (sem.\ $+$ struct.) & 0.5816 & 0.0582 & 0.1437 \\
Mode scoring (batched)              & 0.0013 & 0.0001 & ---    \\
End-to-end per workflow             & ---    & 0.0808 & 0.2282 \\
\midrule
\multicolumn{4}{l}{\emph{End-to-end scalability by workflow size}} \\
Small  ($12$ nodes)                 & ---    & 0.0243 & 0.0540 \\
Medium ($21$--$27$ nodes)           & ---    & 0.0599 & 0.0681 \\
Large  ($30$--$121$ nodes)          & ---    & 0.1388 & 0.3114 \\
\midrule
\multicolumn{4}{l}{\emph{Online incremental scoring (per-step latency)}} \\
Step $1$                            & ---    & 0.0048 & 0.0079 \\
Step $10$                           & ---    & 0.0302 & 0.1237 \\
Step $20$                           & ---    & 0.1048 & 0.5727 \\
Step $80$                           & ---    & 0.0488 & 0.0769 \\
\bottomrule
\end{tabular}
\caption{Runtime overhead of \sysname: offline per-workflow cost, scalability with workflow size, and online incremental per-step latency.}
\vspace{-0.8cm}
\label{tab:overhead}
\end{table}

Table~\ref{tab:overhead} characterizes deployment overhead. Offline per-workflow latency, dominated by graph construction and dual-encoder inference, remains well below one second even for the largest workflows, while mode scoring is negligible. Online per-step latency stays in the tens of milliseconds at early and late prefixes (mean $4.8$--$48.8$ ms at steps $1$ and $80$), rising to roughly $0.1$ s mean ($\sim 0.57$ s P95) at intermediate steps as graph reconstruction grows, before amortizing downward. Thus, \sysname supports both trailing per-workflow and interactive per-step monitoring without stalling the agent pipeline, meeting the runtime requirements for always-on monitoring at the evaluated scale.

\section{Related Work}
\label{sec:related}

\textbf{Agentic AI security.}
Existing safeguards for agentic AI fall into three broad categories. System-level isolation approaches such as IsolateGPT~\cite{wu2025isolategpt} and ACE~\cite{li2025ace} decouple execution environments to limit the blast radius of compromised agents. Policy- and analysis-based defenses such as AgentArmor~\cite{wang2025agentarmor}, SafeFlow~\cite{li2025safeflow}, and ShieldAgent~\cite{chenshieldagent} enforce access controls or verify agent actions against predefined safety rules. Runtime monitoring systems such as TraceAegis \cite{liu2025traceaegis}, DRIFT~\cite{li2025drift}, GUARDIAN~\cite{zhou2025guardian}, and MindGuard~\cite{wang2025mindguard} analyze execution traces or interaction graphs to flag specific threat or coordination patterns. These mechanisms provide important protections at well-defined layers of the agentic stack, but each typically targets a specific violation type or relies on hand-crafted rules; \sysname is complementary, providing a benign-only, attack-agnostic detection layer that surfaces both adversarial manipulations and inherent failures from observed workflows without relying on attack signatures or behavioral specifications.

\textbf{Graph-based anomaly detection.}
Graph anomaly detection has been studied extensively in complex systems. Reconstruction-based methods such as F-GAE~\cite{feng2022fgae} flag structural deviations through autoencoder reconstruction error. Contrastive learning approaches such as CoLA~\cite{liu2022cola}, AD-GCL~\cite{suresh2021AD-GCL}, and TopoGCL~\cite{chen2024topogcl} learn anomaly-sensitive graph representations from unlabeled data, while system-log defenses such as MAGIC~\cite{jia2024magic}, SLOT~\cite{qiao2025slot}, and TRAIL~\cite{lee2025trail} apply graph learning to advanced persistent threat detection. These methods provide strong generic graph signal but are not tailored to agentic execution semantics; \sysname adapts the contrastive paradigm with workflow-aware invariants and counterfactual hard negatives, and pairs it with multi-prototype inference that respects the multimodal nature of benign agentic workflows.

\textbf{LLM safety and benchmarks.}
LLM safety research has focused largely on prompt-level vulnerabilities and alignment, including prompt injection attacks~\cite{greshake2023not, zhan2024injecagent} and defenses based on filtering or robust training~\cite{debenedetti2025defeating, jin2026af}. Benchmarks such as Agent Security Bench~\cite{zhang2025agent_security_bench}, AgentHarm~\cite{andriushchenko2024agentharm}, Agent-SafetyBench~\cite{zhang2024agent}, AgentErrorBench~\cite{zhu2025llmagenterror}, and ATBench~\cite{li2026atbench} provide standardized evaluations for diverse agent vulnerabilities and failure modes. We use the latter three as external testbeds for \sysname, spanning explicit adversarial manipulations, non-adversarial execution failures, and broad safety-in-the-wild settings.

\section{Conclusion}
\label{sec:conclusion}

We presented \sysname, a workflow-level anomaly detection framework for agentic AI systems. Integrated into the runtime observation layer, \sysname consolidates multi-agent execution traces into directed workflow graphs capturing semantic context and structural dependencies. Trained only on benign workflows, it models normality as a multi-mode benign manifold and detects adversarial manipulations, intrinsic failures, and zero-day anomalies under a unified decision rule. Across three public agentic benchmarks, \sysname sustains high recall at a sub-$1\%$ false positive rate, with per-workflow and per-step latencies suitable for online monitoring.

\section*{Acknowledgements}

This work was supported in part by the Office of Naval Research under grant N00014-24-1-2663, N00014-24-1-2730, the National Science Foundation under grants 2154930, 2312447, 2247560, 2332675, 2433904, and 2235232, and the Virginia Commonwealth Cyber Initiative (CCI).

\bibliographystyle{ACM-Reference-Format}
\bibliography{custom}

\end{document}